\documentclass[preprint,amsmath,amssymb]{revtex4}
\usepackage{graphicx}
\usepackage{tabularx}
\usepackage{makecell}
\usepackage{dcolumn}
\usepackage{adjustbox}
\usepackage{hyperref}
\usepackage{bm}

\begin{document}

\title{Embedding diagrams in stationary spacetimes}
\author{${\rm H. \;Sadegh}$ \footnote{Electronic
address:~hamed.sadegh@ut.ac.ir}, ${\rm E.\;Kiani}$ \footnote {Electronic address:~ehsan.kiani@ut.ac.ir} and ${\rm M.\;Nouri-Zonoz}$\footnote{Electronic
address:~nouri@ut.ac.ir\; (Corresponding author)} }
\affiliation{Department of Physics, University of Tehran, North Karegar Ave., Tehran 14395-547, Iran.}

\begin{abstract}
 We find the spatial and dynamic embedding diagrams in some  stationary spacetimes. The spatial embeddings include the  NUT, pure NUT and Kerr spacetimes. In the case of pure NUT spacetime, the spatial embedding equations are solved in terms of the elliptic integrals. In other cases we obtain the spatial embedding diagrams by numerical integration of the corresponding embedding equations. These embedding diagrams are then compared by calculating their Gaussian and mean curvatures. We also find the dynamic embedding diagrams of NUT and pure NUT spacetimes.
\end{abstract}
\maketitle
%%%%%%%%%%%%%%%%%%%%%%%%%%%%%%%%%%%%%%%%%%%%%%%%%%%%%%%%%%%%%%
\section{Introduction}
To better understand and visualize the underlying  geometry in a curved spacetime one could find and employ its 2-dimensional embedding diagrams.
The well-known Flamm's paraboloid \cite{Flamm} gives a snapshot of the equatorial Schwarzschild spacetime embedded in a 3-dimensional Euclidean space. This helps to visualize both the {\it spatial} geometry represented by this solution, and its time-projected geodesics (particle orbits). There are also the dynamic version of the embedding diagrams introduced in \cite{Marolf}, which embed the $(t,r)$-surface (radial plane) of a curved spacetime in a 3-dimensional Minkowski spacetime.\\
Here we will discuss both types of the embeddings, the spatial and the dynamic ones corresponding to the  equatorial surfaces, and radial planes in stationary spacetimes respectively. Obviously the embedding diagrams of sytationary spacetimes are expected to be different from those given for static spacetimes. Two vacuum stationary spacetimes which are somewhat direct generalizations of the Schwarzschild solution are the Kerr \cite{Kerr}, and NUT \cite{NUT} spacetimes. Both are axially symmetric spacetime with two parameters, one of them the mass parameter. The Kerr solution interpreted as the spacetime of a rotating mass includes the angular momentum per unit mass as the second parameter, while the NUT solution which is interpreted as the gravitational analog of a magnetic monopole, or the so called {\it gravitomagnetic monopole}, includes the NUT parameter also called the {\it  magnetic mass}, as the second parameter.\\
The well-known Kerr solution in Boyer-Lindquist coordinates \cite{BL} is given by the metric,
\begin{equation}\label{n0}
d{s^2} = (1 - \frac{{2mr}}{{{\rho ^2}}})d{t^2} + \frac{{4mar{{\sin }^2}\theta }}{{{\rho ^2}}}dtd\phi  - \frac{{{\rho ^2}}}{\Delta }d{r^2} - {\rho ^2}d{\theta ^2} - ({r^2} + {a^2} +
\frac{{2m{a^2}r{{\sin }^2}\theta }}{{{\rho ^2}}}){\sin ^2}\theta d{\phi ^2}
\end{equation}
with
\begin{equation}
{\rho ^2} = {r^2} + {a^2}{{\cos }^2}\theta \;\;\;, \;\;\; \Delta  = {r^2} - 2mr + {a^2}.
\end{equation}
in which $m$ and $a$ are the mass and angular momentum per unit mass respectively. For $a=0$ it reduces to Schwarzschild solution, and for $m=0$, and $r\ge0$ to flat spacetime in oblate spheroidal coordinates \footnote{For a recent discussion on the zero mass limit of the extended ($-\infty < r < \infty$) Kerr spacetime refer to \cite{Gibb}.}. For $m^2 > a^2$ it has two horizons (outer and inner) with radii
\begin{eqnarray}\label{KH}
 r_{\pm} = m \pm \sqrt{m^2-a^2}
\end{eqnarray}
and a ring singularity of radius $a$ lying in the equatorial plane behind its horizons \cite{Carter}.\\
NUT spacetime on the other hand has a number of exotic, and at the same time interesting  properties. Its metric in Schwarzschild-type coordinate is given by
\begin{equation}\label{n1}
ds^2 = f(r) (dt - 2l \cos\theta d\phi)^2 - \frac{dr^2}{f(r)} - (r^2 + l^2) d\Omega^2
\end{equation}
with
\begin{equation}\label{f}
f(r) = \frac{r^2 - l^2 - 2mr}{r^2 + l^2},
\end{equation}
in which $m$ and $l$ are the mass, and NUT parameters respectively. For $l=0$ we recover the Schwarzschild metric, whereas for $m=0$, unlike the Kerr case, we do not find flat spacetime in an exotic coordinate, but the so called pure NUT spacetime which is a one-parameter stationary spcetime. Taking the radial coordinate $0 < r < \infty $, there is a coordinate singularity at $r_H = m  + (m^2 + l^2)^{1/2}$ (where $f(r)=0$) representing the NUT horizon, which unlike the horizons of  Schwarzschild and Kerr black holes, is not hiding an intrinsic singularity.\\
Along the z-axis ($\theta=0, \pi$ ) it has a string singularity \cite{Misner} which in comparison with a Dirac monopole in electromagnetism, and its string singularity,  allows an interpretation of NUT spacetime as the spacetime of a mass endowed with a gravitomagnetic monopole charge (the NUT parameter $l$) \cite{DN}. This justifies its {\it physical} spherical symmetry despite its mathematically axisymmetric appearance \cite{Zimm}-\cite{DLBMNZ}. Through a coordinate transformation, Bonnor reduced the singularity to half-axis, giving the NUT spacetime another interpretation in terms of a central mass plus a semi-infinite massless source of angular momentum \cite{Bonnor}. Another interesting feature of NUT spacetime, also originated from its Dirac-type string singularity, is the fact that it is {\it locally} but not globally asymptotically flat. To put it another way, its Riemann tensor  vanishes as $r\rightarrow \infty$, but its metric reduces to that of flat spacetime except in the direction of the singularity. Observationally, the NUT spacetime, as the spacetime around an astrophysical object, could be identified from its distinct effects on light rays as a gravitational lens \cite{MNDL, SRMN}. \\
In what follows we will begin with spatial embedding diagram of the equatorial NUT spacetime in three dimensional Euclidean space. This can not be done analytically, so by reducing the embedding equation to the cases of Schwarzschild and pure NUT spaces, we show that one can find, in both cases, the embedding diagrams both analytically and numerically which agree perfectly. This provides an evidence for the reliability of our numerical calculations for this type of embedding diagrams for Kerr and full NUT spacetimes. Next we calculate the Gaussian and mean curvatures of the embedding surfaces, and to better compare all these embedding diagrams, and their curvatures as a function of the {\it embedding radial coordinate}, we  plot them  in a single diagram by fixing their corresponding horizons at a given common value.\\
Finally we will discuss the dynamic embedding of NUT spacetime as a prototype of this kind of embeddings for stationary spacetimes, and compare its diagram with the corresponding diagram given for the Schwarzschild spacetime.
%%%%%%%%%%%%%%%%%%%%%%%%%%%%%%%%%%%%%%%%%%%%%%%%%
\section{Embedding diagram of equatorial NUT spacetime}
A time section ($t=constant$) of the NUT spacetime in the equatorial plane ($\theta= \frac{\pi}{2}$) is given by the following 2-dimensional metric.
\begin{equation}\label{n2}
ds^2 = \frac{dr^2}{f(r)} + (r^2 + l^2) d\phi^2
\end{equation}
Obviously, taking the string singularity along the $z$-axis, this is a 2-dimensional asymptotically flat space.
As in the case of Schwarzschild spacetime, to better visualize the geometry of the above 2-surface we need to embed it into a three dimensional Euclidean space. To do so first we take
$(r^2 + l^2) = R^2$, so that the above metric reduces to
\begin{equation}\label{n3}
ds^2 = \frac{R^4}{(R^2 - l^2)(R^2 - 2l^2 - 2m(R^2 - l^2)^{1/2})} dR^2 + R^2 d\phi^2
\end{equation}
Now using the standard procedure we can embed this surface in a  3-dimensional Euclidean space with the following metric
\begin{equation}\label{n4}
ds^2 =dZ^2 + dR^2 + R^2 d\phi^2
\end{equation}
written in the following form
\begin{equation}\label{n5}
ds^2 =(1 + \frac {dZ^2}{dR^2})dR^2 + R^2 d\phi^2.
\end{equation}
So that comparing the above equation with \eqref{n3}, we end up with the following integral for $Z=Z(R)$
\begin{equation}\label{d1}
Z(R) = \int \left(\frac{3l^2 R^2 + 2m(R^2-l^2) ^{3/2} -2l^4}{(R^2-l^2)[R^2-2l^2 - 2m(R^2 - l^2)^{1/2}]}\right)^{1/2} dR.
\end{equation}
in which the result of integration is valid for $R > R_H =({r_H}^2 + l^2)^{1/2} $. Now there are different cases that one can consider:\\
{\bf I)} The case of Schwarzschild spacetime ($l=0$) : In this well known case, where $R=r$, the above integral leads to the following curve
\begin{equation}\label{F1}
Z^2 =8m(r-2m)
\end{equation}
whose surface of revolution about the Z-axis, valid for $r > r_H = 2m$, is the well-known Flamm's paraboloid. In Figure 1 the above anlytical curve and  the numerical integration of \eqref{d1} for $l=0$ are shown on the same plot which agree perfectly. This could be taken as a test of validity and precision of the employed numerical integration.\\
%%%%%%%%%%%%%%%%%%%%%%%%%%%%%%
{\bf II)} The case of pure NUT spacetime ($m=0$): Here we have the following integral valid for $R > R_H = \sqrt{2}l$,
\begin{equation}\label{d11}
Z(R) = \int \left(\frac{3l^2 R^2 - 2l^4}{(R^2-l^2)(R^2-2l^2)}\right)^{1/2} dR.
\end{equation}
%%%%%%%%%%%%%%%%%%%%%%%%%%%%%%%%%%%%%%%%%%%%%%%%%%%%%%%%%%%%
\begin{figure}\label{SCH}
\includegraphics[scale=1.1]{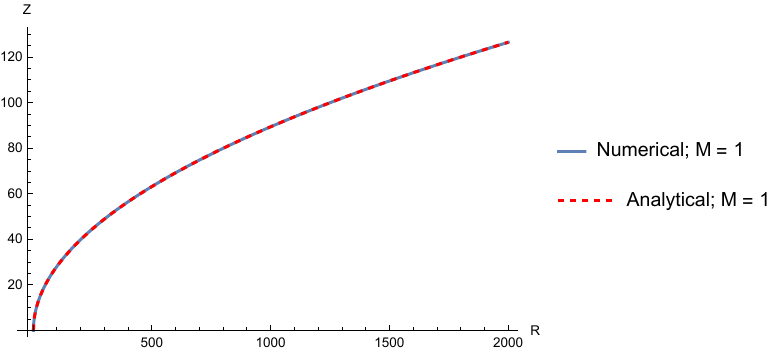}
\caption{Flamm's parabola in Schwarzscild spacetime with $m=1$.}
\end{figure}
%%%%%%%%%%%%%%%%%%%%%%%%%%%%%%%%%%%%%%%%%%%%%%%%%%%%%%%%%%%%
\begin{figure}\label{SCH3D}
\includegraphics[scale=1.2]{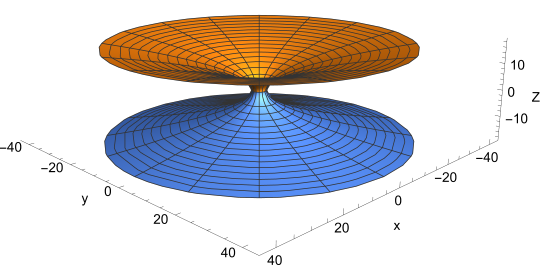}
\caption{Flamm's paraboloid in Schwarzschild spacetime.}
\end{figure}
%%%%%%%%%%%%%%%%%%%%%%%%%%%%%%%5
\begin{figure}\label{PNUT}
\includegraphics[scale=1.1]{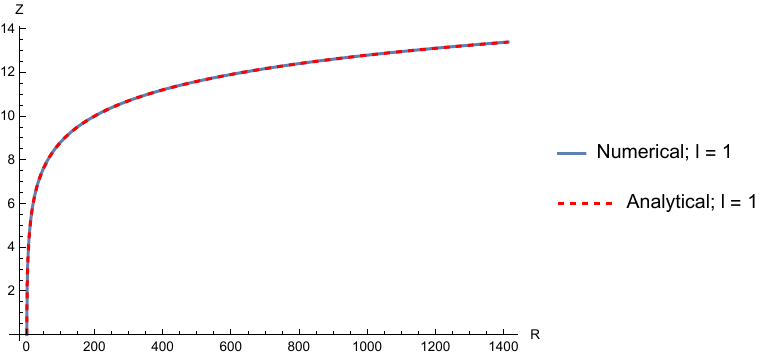}
\caption{Analog of Flamm's parabola in pure NUT spacetime with $l=1$}
\end{figure}
%%%%%%%%%%%%%%%%%%%%%%%%%%%%%%%%%%%%%%%%%
\begin{figure}\label{PNUT3D}
\includegraphics[scale=1.2]{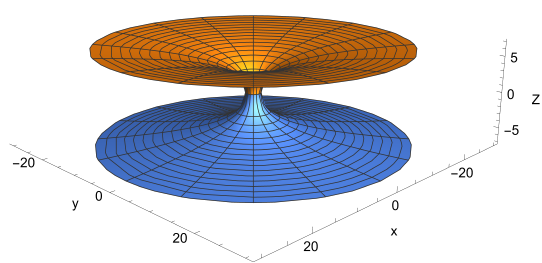}
\caption{Analog of Flamm's paraboloid in pure NUT spacetime with $l=1$.}
\end{figure}
%%%%%%%%%%%%%%%%%%%%%%%%%%%%%%%%%
which interestingly enough could be calculated in terms of elliptic integrals through the change of variable
$ u = -2 l^4-3 l^2 R^2 $,
leading to the following curve
\begin{eqnarray}\label{Pure NUT analytic}
Z(R) =\pm \frac{1}{2} l \left(F\left(\csc ^{-1}\left(\sqrt{\frac{2 l^2}{R^2-2 l^2}+2}\right)|\frac{1}{2}\right)+3 \Pi \left(2;\csc ^{-1}\left(\sqrt{\frac{2 l^2}{R^2-2 l^2}+2}\right)|\frac{1}{2}\right)\right)
\end{eqnarray}
where
\begin{eqnarray}\label{incomplete elliptic integral}
\Pi (n;\phi |m)=\int _0^{\phi }d \theta \frac{1}{\sqrt{1-m \sin ^2(\theta )} \left(1-n \sin ^2(\theta )\right)}
\end{eqnarray}
is the incomplete elliptic integral of the third kind, and
\begin{eqnarray}\label{elliptic integral of the first kind}
F(\phi |m)=\int_0^{\phi } \frac{1}{\sqrt{1-m \sin ^2(\theta )}} \, d\theta
\end{eqnarray}
%%%%%%%%%%%%%%%%%%%%%%%%%%%%%%%%%%%
is the incomplete elliptic integral of the first kind. Having the values of these integrals, the embedding curve of pure NUT is depicted, along with the direct numerical integration, in Fig. 3. Again the two results coincide perfectly. The surface revolution of the curves in Figures 1 and 3, corresponding to Flamm's paraboloid and its analog in pure NUT spacetime, are shown in Figures 2 and 4 respectively. From figures  1 and 3 it is obvious that the embedding geometry in the case of pure NUT spacetime, as  compared to Flamm's parabola, starts flattening at shorter embedding radii.\\
%%%%%%%%%%%%%%%%%%%%%%%%%
\begin{figure}\label{NUTC}
\includegraphics[scale=0.6]{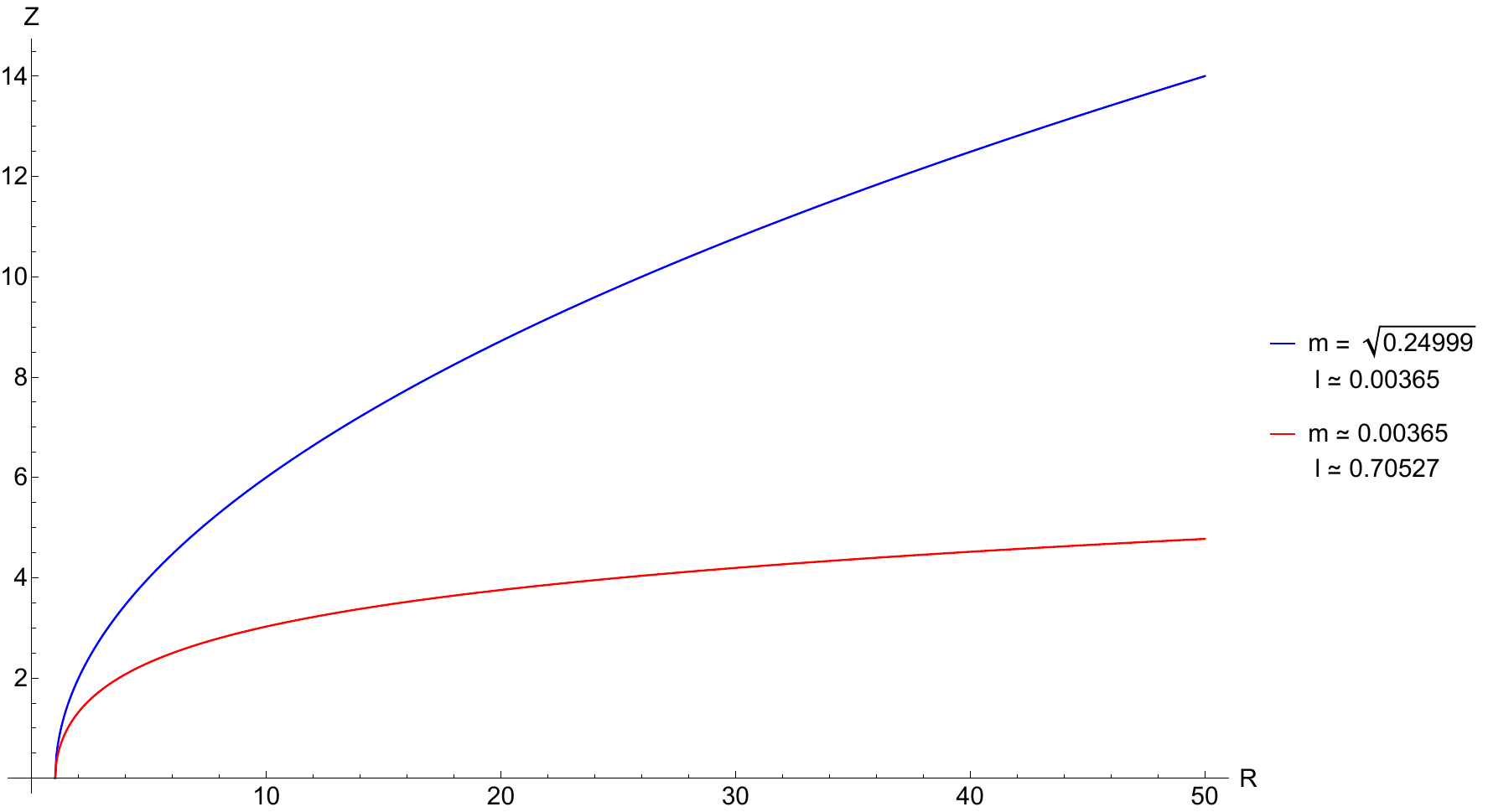}
\caption{Embedding diagram of NUT spacetime with two different sets of values for $m$ and $l$, but the same embedding horizon radius $R_H = 1$.}
\end{figure}
%%%%%%%%%%%%%%%%%%%%%%%%%%%%%%%%%
{\bf III)}  The general case of NUT spacetime with non-zero mass and NUT parameter: In this case the integral \eqref{d1} can not be performed analytically or in terms of known functions such as elliptic integrals, and so we employ numerical integration to obtain the embedding diagram. For example the resulted embedding curves for two different sets of values for $m$ and  $l$ with the same embedding horizon radius $R_H = 1$, are shown in Fig. 5. Comparing with Figures 1 and 3, this plot shows how the larger values of $m$ or $l$ dominate the asymptotic behaviour of the embedding curve in approaching to those corresponding to Schwarzschild and pure NUT spacetimes respectively. Below we will see that this same pattern is repeated in  the behaviour of the  Gaussian and mean curvatures of the NUT embedding diagrams. \\
Before doing so it should be  pointed out that in a recent study \cite{PFRN} metamaterial analog of closed photon orbits, the so called {\it photon rings}, in the equatorial NUT spacetime are simulated. These exact ray-tracing simulations are based on the metamaterial's index of refraction which is adapted from the corresponding spacetime index of refraction \cite{NPF}. Having the above embedding diagrams of NUT spacetime helps us to visualize these photon rings as well as other null and timelike orbits in NUT spacetime. Indeed they are the time-projected  photon orbits on the above introduced embedding diagram. As in the case of Schwarzschild spacetime \cite{Rindler}, it is expected that the spatial curvature of NUT spacetime, manifested in its embedding diagram, to be partially responsible for light bending around a NUT hole \cite{MNDL}.
%%%%%%%%%%%%%%%%%%%%%%%%%%%%%%%%%%%%%%%%%%%%%%%%%%%%%%%%%%%%%%%%
\subsection{Gaussian and mean curvatures of NUT embedding diagrams}
Another way that one could compare the NUT, Schwarzschild and pure NUT spacetimes embedding diagrams, is through their Gaussian and mean curvatures. Obviously since these two dimensional surfaces only depend on the embedding radial coordinate $R$, so will be their Gaussian and mean curvatures. The general formulae for the Gaussian and mean curvatures of these surfaces are given by \cite{Frankel}
\begin{equation}\label{GC}
K = \frac{Z' Z''}{R \left(1+Z'^2\right)^2}
\end{equation}
\begin{equation}\label{MC}
H = \frac{R Z''+Z'^3+Z'}{2 R \left(1+Z'^2\right)^{3/2}}
\end{equation}
where $ `` \prime `` $ denotes differentiation with respect to the embedding radial coordinate $R$. For the NUT family one finds
\begin{eqnarray}\label{GKNUT}
K_{NUT} = \frac{4 l^4+l^2 \left(4 m \sqrt{R^2-l^2}-3 R^2\right)-m R^2 \sqrt{R^2-l^2}}{R^6}
\end{eqnarray}

\begin{eqnarray}\label{MCNUT}
H_{NUT} = \frac{2 l^4+\sqrt{R^2-l^2} \left(2 l^2 m+m R^2\right)}{2 R^3 \sqrt{l^4+2 m \left(R^2-l^2\right)^{3/2}+3 l^2 \left(R^2-l^2\right)}}
\end{eqnarray}
which for $l=0$, and $m=0$ reduce to the following values
\begin{equation}\label{GKSCH}
 K_{Sch} = -\frac{m}{R^3}\;\;\;\;\;\;\; ; \;\;\;\;\;\;\;\; H_{Sch} = \frac{1}{2R}\sqrt{\frac{m}{2R}}
\end{equation}
and
\begin{eqnarray}\label{GKPNUT}
K_{PNUT} = \frac{4 l^4-3 l^2 R^2}{R^6}\;\;\;\;\;\;\; ; \;\;\;\;\;\;\;\; H_{PNUT} =\frac{l^3}{R^3 \sqrt{3 R^2-2 l^2}}
\end{eqnarray}
respectively. Obviously the NUT spacetime embedding curvatures \eqref{GKNUT} and \eqref{MCNUT} reduce to their limiting values \eqref{GKSCH} and \eqref{GKPNUT}, when $\frac{l}{R} \ll 1$ and $\frac{m}{R} \ll 1$, respectively. Indeed these are the same limiting values shown in the two diagrams in Fig. 5.
%%%%%%%%%%%%%%%%%%%%%%%%%%%%%%%%%%%%%%%%%%%%%%%%%%%%%%%%%%%%%%%%
\section{Embedding diagrams of Reissner-N\"{o}rdstrom and Kerr spacetimes}
Although Reissner-N\"{o}rdstrom (R-N) spacetime is a static spacetime, but having two parameters makes it a good example to compare its embedding diagram with that of NUT spacetime we examined in the last section, and the Kerr spacetime which we will discuss later on in this section.
%%%%%%%%%%%%%%%%%%%%%%%%%%%%%%%%%%%%%%%%%%%%%%%
\subsection{Embedding diagram of equatorial R-N spacetime}
Embedding diagrams for R-N spacetime, both spatial and dynamic, have already discussed in the literature (see \cite{Stuch1}-\cite{Dadhich}).
The R-N spacetime as a static spherically symmetric solution of the Einstein-Maxwell equations, is given by the following line element
\begin{eqnarray}\label{RN metric}
ds^{2} = f(r)dt^2 - \frac{dr^2}{f(r)} - r^2d\Omega^2\\
f(r) =  1 - \frac{2m}{r} + \frac{q^2}{r^2}
\end{eqnarray}
in which the two parameters are mass $m$ and charge $q$. It reduces to Schwarzschild spacetime for $q=0$, and has two horizons at $r_{\pm} = m \pm \sqrt{m^2-q^2}$. At a given  instant of time and in the equatorial plane we have
\begin{eqnarray}\label{RN in plane}
ds^{2} = \frac{dr^2}{f(r)}+r^2d\phi^2.
\end{eqnarray}
Noting that here, as in the case of Schwarzschild spacetime, $R=r$, and  employing the standard embedding procedure, the equatorial R-N embedded in Euclidean space is given by the curve $Z(R)\equiv Z(r)$ through the following integral,
\begin{eqnarray}\label{RN z equation}
Z(r) = \int \sqrt{\frac{2mr - q^2}{r^2 - 2mr + q^2}}dr
\end{eqnarray}
valid for $r > r_+$, leading to
\begin{eqnarray}\label{rnz1}
Z(r)=\frac{2 \left(E\left(\frac{4 m \sqrt{m^2-q^2}}{q^2+2 m \left(\sqrt{m^2-q^2}-m\right)}\right)-E\left(\csc ^{-1}\left(\frac{\sqrt{2} \sqrt[4]{m^2-q^2}}{\sqrt{-m+r+\sqrt{m^2-q^2}}}\right)|\frac{4 m \sqrt{m^2-q^2}}{q^2+2 m \left(\sqrt{m^2-q^2}-m\right)}\right)\right)}{\sqrt{\frac{1}{2 m \left(\sqrt{m^2-q^2}-m\right)+q^2}}}
\end{eqnarray}
where the elliptic integral $E$ is defined as follows
\begin{eqnarray}\label{E}
E(\phi|m)=\int^\phi_0 (1-m\sin^2\theta)^{\frac{1}{2}} d\theta \;\;\; ; \;\;\; E(m)\equiv E(\pi/2|m).
\end{eqnarray}
The embedding diagram of R-N spacetime, based on \eqref{rnz1} is shown in Fig. 6.
%%%%%%%%%%%%%%%%%%%%%%%%%%%%%%%%%%%%%%%%%%%55
\begin{figure}\label{RN}
\includegraphics[scale=0.55]{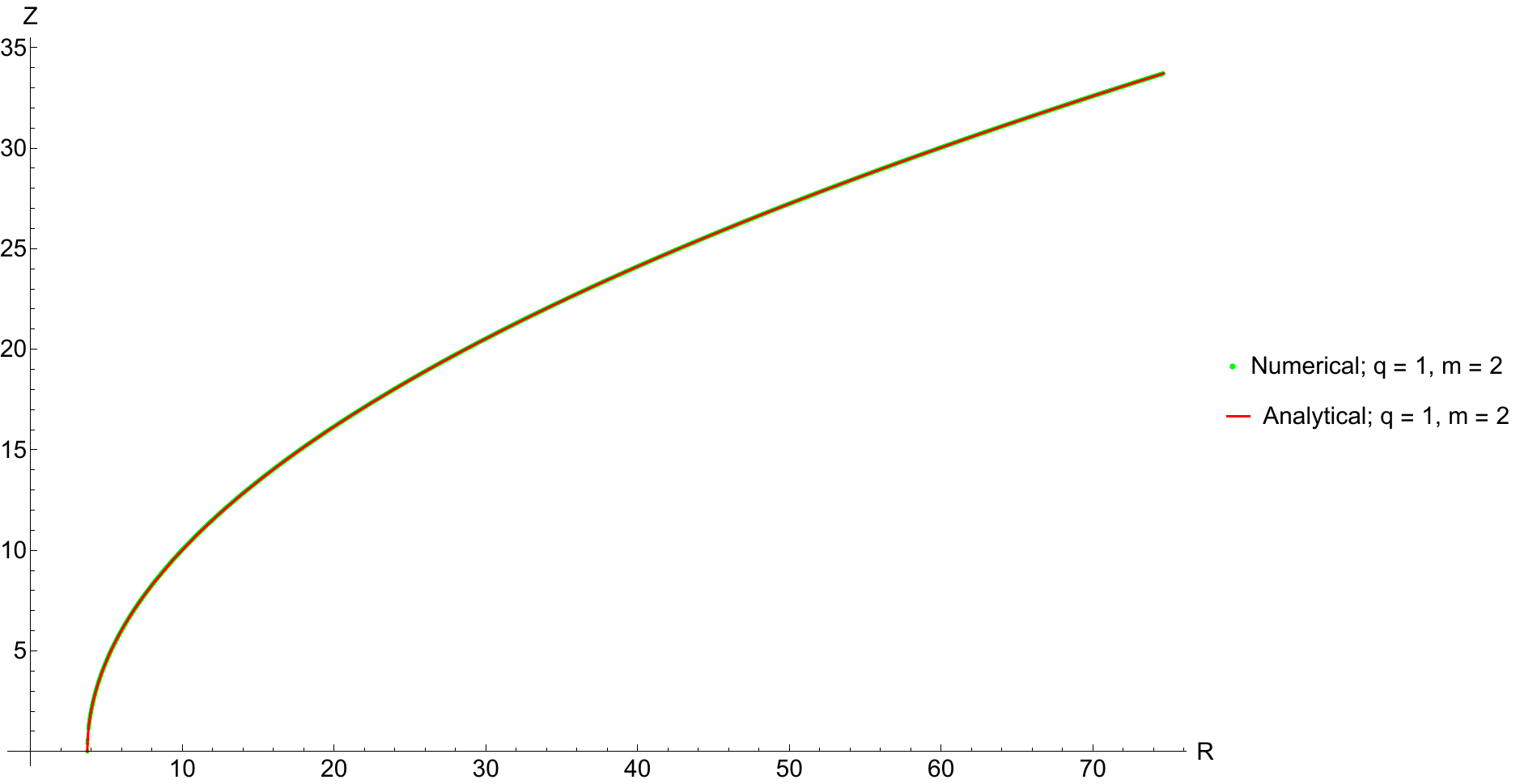}
\caption{Analog of Flamm's parabola for R-N spacetime with $m=2$ and $q=1$.}
\end{figure}
%%%%%%%%%%%%%%%%%%%%%%%%%%%%
Using the formulas \eqref{GC} and \eqref{MC}, the Gaussian and mean curvatures of embedding diagrams of R-N are given by
\begin{eqnarray}\label{RN GC}
K_{R-N} = \frac{q^2-m R}{R^4}\;\;\;\; ; \;\;\;\; H_{R-N} = \frac{m}{2 R \sqrt{2 m R-q^2}}
\end{eqnarray}
%%%%%%%%%%%%%%%%%%%%%%%%%%%%%%%%%%%%%
\subsection{Embedding diagram of equatorial Kerr metric}
Embedding diagrams for equatorial kerr metric were first discussed in \cite{Bardeen}, in which the authors gave a schematic illustration of these diagrams for near-extreme Kerr black holes.  Also embedding diagrams in Kerr-Newmann optical reference geometry are discussed in \cite{Stuch2}.\\
Kerr metric in the equatorial plane, and at an instant of time reduces to
\begin{eqnarray}\label{Kerr in plane}
ds^{2} = \frac{r^2}{r^2-2mr+a^2}dr^2+(r^2+a^2+\frac{2ma^2}{r})d\phi^2.
\end{eqnarray}
Now taking $R^2 = r^2+a^2+\frac{2ma^2}{r}$, from \eqref{KH}, the outer horizon will be at $R_H = 2m$. In other
words $R_H$ in the Kerr case is the same as in the  Schwarzschild metric, so unlike the previous cases there is no trace of the second parameter, (the angular momentum $a$) in the horizon radius given in terms of the embedding radial coordinate.\\
To find $r(R)$ we need to solve the cubic equation $r^3+r(a^2-R^2) + 2ma^2 =0 $ with discriminant $D= (\frac{a^2-R^2}{3})^{3} + m^2a^4$. Taking into account that $R \geq2m$, and $a\leq m$, the largest value of $D$, which is {\it zero}, is given by  $R = 2m$, and $a=m$. In other words we always have $D\leq 0$ with the equal sign holding for an extreme Kerr black hole. In the general case  $D<0$, one can show that the only solution which reduces to $r=R$ for $a=0$ is given by
\begin{eqnarray}\label{Kerr R}
r= \frac{2}{\sqrt{3}}  \sqrt{R^2-a^2} \cos \left(\frac{1}{3} \cos ^{-1}\left(-\frac{3 \sqrt{3} a^2 m}{\left(R^2-a^2\right)^{3/2}}\right)\right).
\end{eqnarray}
Now applying the standard  embedding procedure we end up with the following equation for $Z(R)$ to be solved for the embedding diagram
\begin{eqnarray}\label{KD}
\frac{dZ}{dR}= \sqrt{\frac{m \left(-a^6 m+a^4 r \left(2 m^2-m r+2 r^2\right)+4 a^2 r^4 (r-m)+2 r^7\right)}{\left(r^3-a^2 m\right)^2 \left(a^2+r (r-2 m)\right)}}
\end{eqnarray}
The results of numerical integration of \eqref{KD}, valid for
$R > R_H = 2m$, gives the embedding diagrams in Fig. 7. These are for two different sets of values for $m$ and $a$, with the same $R_H = 1$. It is seen that when the value of $a$ is very close to $m$, as in the case of extreme Kerr black hole, the embedding curve tends to become flat with a less steeper slope.
 %%%%%%%%%%%%%%%%%%%%%%%%%%%%%%%%%%%%%%%%%%%55
\begin{figure}\label{Kerr}
\includegraphics[scale=0.57]{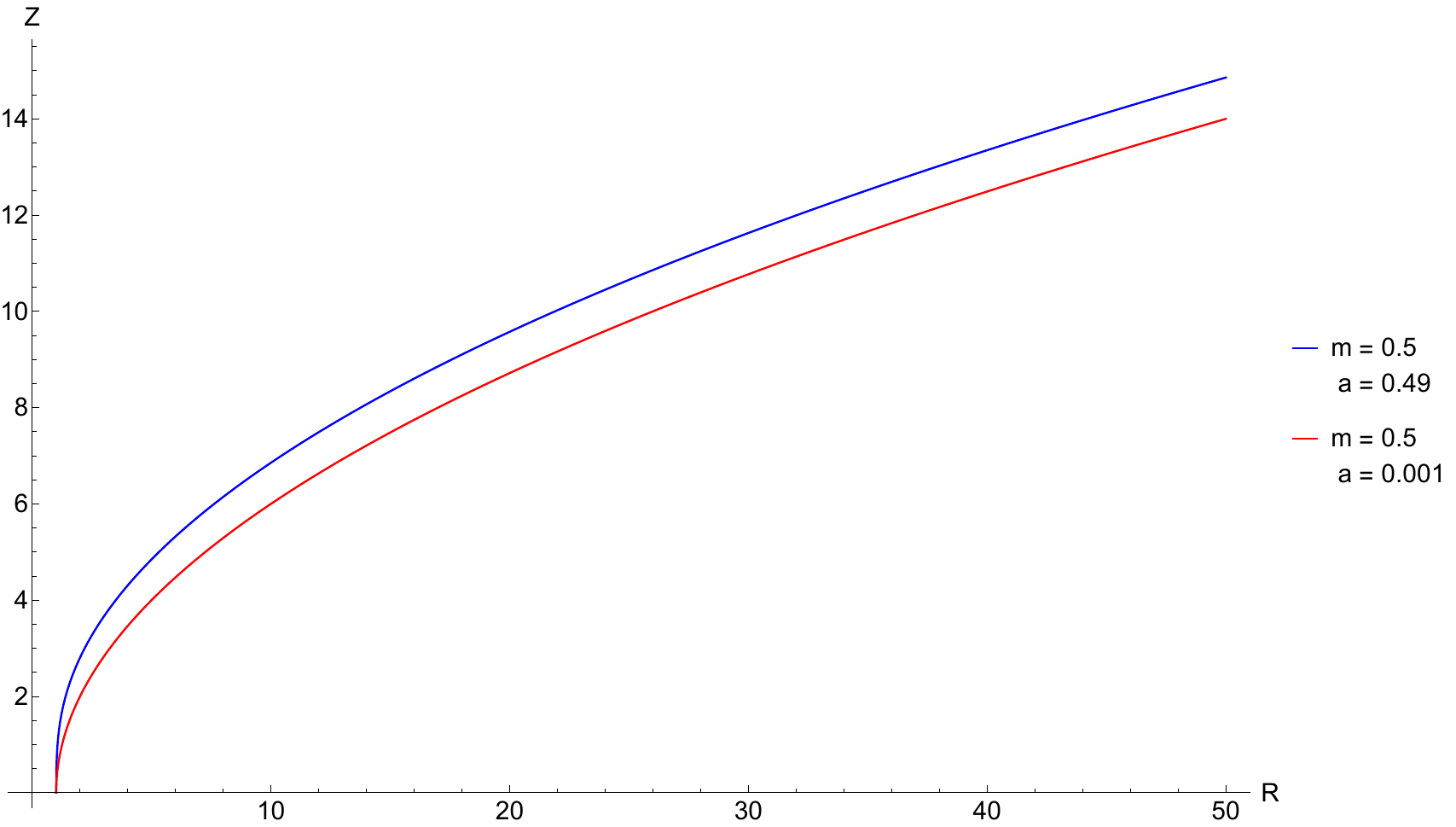}
\caption{Embedding curve for Kerr spacetime with two different sets of values for $m$ and $a$, but the same embedding horizon radius $R_H = 1$.}
\end{figure}
%%%%%%%%%%%%%%%%%%%%%%%%%%%%%%%%%%%%%
In the case of Kerr spacetime the Gaussian and mean curvatures of the  embedding diagram are more detailed functions given by,
\begin{eqnarray}\label{Kerr GC}
K = \frac{m \left(-a^6 (5 m+3 r)+a^4 r \left(8 m^2+2 m r-7 r^2\right)+a^2 r^4 (11 m-5 r)-r^7\right)}{r^4 \left(a^2 (2 m+r)+r^3\right)^2}
\end{eqnarray}
and
\begin{flalign}
H & \nonumber \\
=&
\frac{1}{2 F} \left(\frac{\Delta}{a^2 r^5 (2 m+r)+r^8}\right)^{3/2}\{ \frac{r^{11/2} \left(G^2 \left(r^3-a^2 m\right)^2-3 A P^4\right)}{\sqrt{\Delta} G^2} \nonumber & \\
-&\frac{A r^{13/2} (m-r) \left(2 a^4 m^2-a^2 m r^3-r^6\right) \left(6 a^4 G m^2+3 a^2 G m r^3\right)}{\Delta^{3/2} G P^6}+\frac{A F^4 \sqrt{r}}{\Delta^{3/2}} \nonumber & \\
+&\frac{A r^{13/2} (m-r) \left(a^2 m-r^3\right) \left(2 a^4 m^2-a^2 m r^3-r^6\right) \sqrt{16 a^4 m^2+10 a^2 m r^3+r^6}}{\Delta^{3/2} G P^5} \nonumber & \\
-& \frac{3 a^2 A m r^{11/2} \left(a^2 m-r^3\right) \left(16 a^4 m^2+10 a^2 m r^3+r^6\right)^{3/2}}{\sqrt{\Delta} G^3 P^5} \nonumber & \\ &+\frac{3 A r^{17/2} \left(-5 a^4 m^2+4 a^2 m r^3+r^6\right)}{\sqrt{\Delta} G^2 P^2} +F^2 \sqrt{\frac{r}{\Delta}} \left(r^3-a^2 m\right)^2 \}\\
A=&a^2 (2 m+r)+r^3 \; ; \;
G=\sqrt{8 a^2 m+r^3} \; ; \; P=\sqrt{2 a^2 m+r^3} \\
F=& \sqrt{\frac{m \left(-a^6 m+a^4 r \left(2 m^2-m r+2 r^2\right)+4 a^2 r^4 (r-m)+2 r^7\right)}{a^2 (2 m+r)+r^3}}
\end{flalign}
As a check point it is easy to see that for $a=0$  the above curvatures reduce to their Schwarzschild values in \eqref{GKSCH}.
%%%%%%%%%%%%%%%%%%%%%%%%%%%%%%%%%%%%%%%%%%%%%%
\begin{figure}\label{AllC}
\includegraphics[scale=0.54]{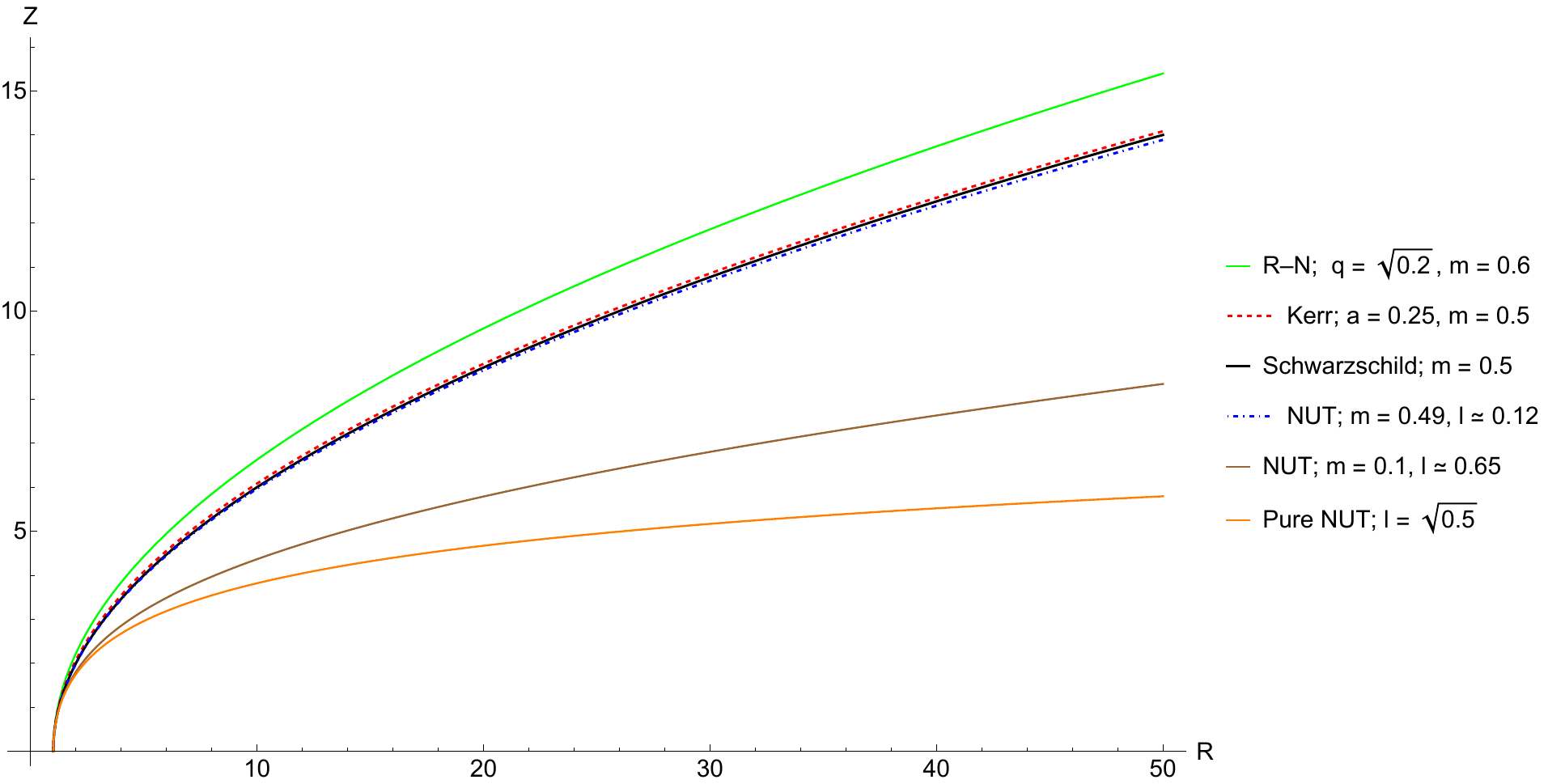}
\caption{Behaviour of the embedded diagrams of different spacetimes, all starting from the horizon radius $R_H =1$.}
\end{figure}
%%%%%%%%%%%%%%%%%%%%%%%%%%%%%%%%%%%
\begin{figure}\label{AllCC}
\includegraphics[scale=0.54]{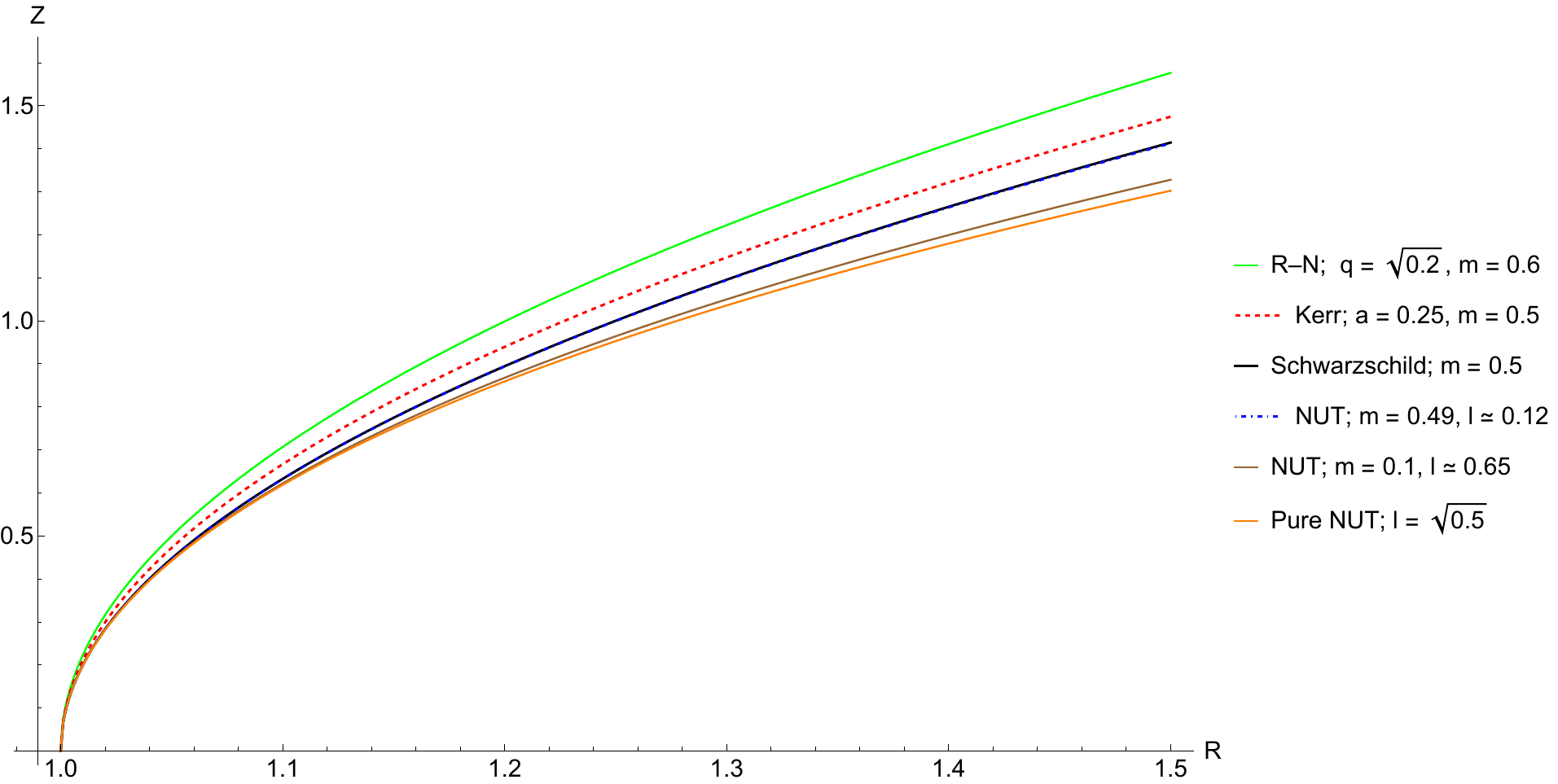}
\caption{Behaviour of the embedding diagrams in the neighbourhood of their common horizon.}
\end{figure}
%%%%%%%%%%%%%%%%%%%%%%%%%%%%%%%%%%%%%%
\begin{figure}\label{GKC}
\includegraphics[scale=1.1]{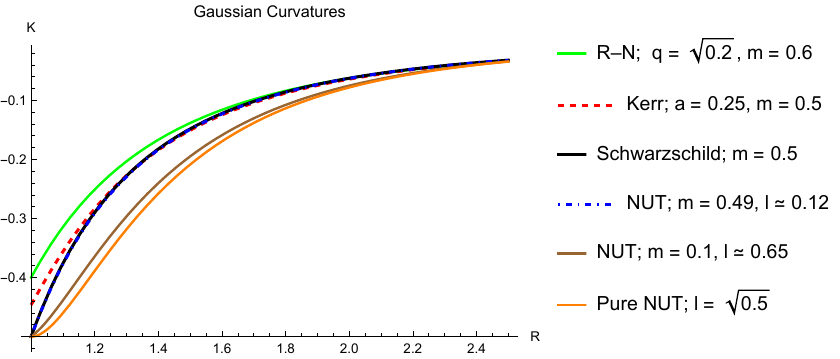}
\caption{Behaviour of the Gaussian curvatures of all the embedding diagrams.}
\end{figure}
%%%%%%%%%%%%%%%%%%%%%%%%%%%%%
\begin{figure}\label{MKC}
\includegraphics[scale=1.1]{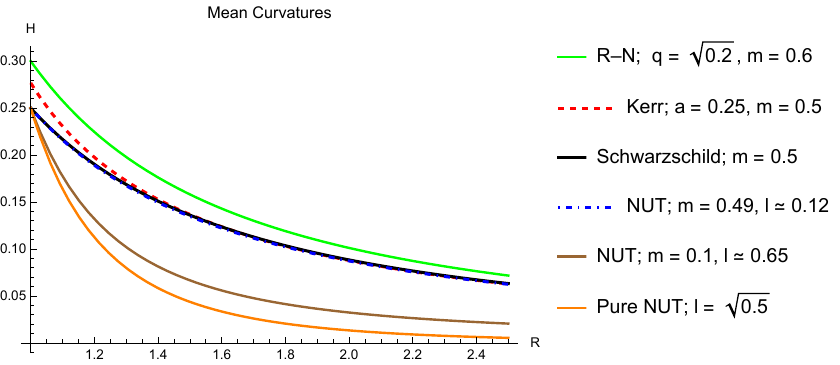}
\caption{Behaviour of the Mean curvatures of all the embedding diagrams.}
\end{figure}
%%%%%%%%%%%%%%%%%%%%%%%%%%%%%%%%%%%%%%%%
\section{Comparison of embedding diagrams and their curvatures}
To compare the effect of different parameters on the embedding diagrams discussed above, and also compare their asymptotic behaviour, one can draw all the embedding curves in the same plot. To this end  we  plot all the curves starting from the same horizon size in terms of the embedding radial coordinate, namely $R_H=1$. As noted previously, by choosing a given value for $R_H$ the corresponding spacetime parameters are either fixed, as in the one-parameter solutions, or are restricted as in the case of two-parameter solutions. The behaviour of the embedding diagrams for all the above-considered spacetimes are given in Figure 8.\\
Since the behavior of embedding diagrams in Schwarzschild, NUT (with $m > l$) and Kerr for equal or almost equal masses look very similar, in Fig. 9  we have magnified the same diagram in the neighbourhood of their common horizon.\\
Gaussian and mean Curvatures of all the embedding diagrams are also plotted against $R$, in Figs. 10 and 11 respectively. Obviously as expected, all the embedding diagrams are open 2-surfaces with negative Gaussian curvatures.
%%%%%%%%%%%%%%%%%%%%%%%%%%%%%%%%%%%%%%%%%%%%%%%%
\section{Dynamic embedding of NUT spacetime}
Unlike spatial embedding diagrams discussed in the previous sections, which lack  information on the spacetime dynamics, the spacetime or {\it dynamic} embedding of a radial plane ($(t,r)$-plane) into the 3-dimensional Minkowski spacetime includes the dynamics of the original spacetime \cite{Marolf}.
In the case of NUT spacetime \eqref{n1}, the radial plane (the $r$-$t$ plane) which is obtained  by setting angular coordinates $\theta$ and $\phi$ equal to constants, is given by
\begin{eqnarray}\label{n11}
ds^2 = f(r) dt^2 - \frac{dr^2}{f(r)}  \\
f(r)=\frac{r^2-l^2-2 m r}{r^2+l^2} \label{n12}.
\end{eqnarray}
This is very similar to the radial plane of any static, spherically symmetric spacetime, including Schwarzschild and Reissner-Nordstrom black holes apart from a different function
$f(r)$ which now  includes both the mass and the NUT parameters. Following \cite{Marolf}, we try to embed this radial plane into the 3-dimensional Minkowski spacetime with the following metric in the Cartesian coordinates;
\begin{eqnarray}
d s^2=d T^2-d X^2-d Y^2.
\end{eqnarray}
To do so first we note that the above metric could be written in cylindrical hyperbolic coordinates (Rindler coordinates), i.e,
\begin{equation}
ds^2 = \rho^2 d \zeta^2-d \rho^2-d Y^2 \label{Rindler}
\end{equation}
by the following coordinate transformations,
\begin{eqnarray}\label{TX1}
%\begin{split}
        \zeta &= \operatorname{tanh}^{-1}\left(\left(\frac{T}{X}\right)^{ \pm 1}\right) \label{hyp} \\
         \rho &= \sqrt{ \pm \left (T^2-X^2\right)}\label{TX2} ,
     %\end{split}
     \end{eqnarray}
where different combinations of $\pm$ signs  gives transformations for different wedges. \\
On the other hand the metric of the local geometry near any smooth non-degenerate horizon, including that of the radial NUT at $r_H$, could be written in a $1+1$ Rindler-like form. So the first step is to write the radial NUT metric in this form. To do so we use the definition of the surface gravity of the generic metric form \eqref{n11}, i.e  $\kappa=\frac{1}{2} \partial_r f(r)|_{r=r_H}$, and the following coordinate transformations
\begin{flalign}\label{Trax}
\zeta&= \kappa t \\
\rho&=\frac{\sqrt{f(r)}}{\kappa} \label{Trax1}
\end{flalign}
which upon applying  to \eqref{n11} lead to the desired form
\begin{equation}
d s^2=\rho^2 d \zeta^2-d \rho^2,
\end{equation}
near the horizon, by taking the leading term in the expansion of $f(r)$ at $r=r_H$. This is the $Y=constant$ surface in metric \eqref{Rindler}, and we need to find a general embedding surface $Y = Y (\zeta,\rho) \equiv Y (t,r) $ to go beyond the above leading approximation. To this end we follow the same procedure employed in \cite{Marolf} for the static spherically symmetric spacetimes. But before doing so we note that our ability to use the above procedure in Schwarzschild spacetime was reflected in the fact that one could use similar hyperbolic transformation as in \eqref{hyp}, to maximally extend Schwarzschild metric, and obtain the Kruskal spacetime which divides the spacetime into four different regions. This is compared to  Minkowski spacetime as the maximal extension of Rindler geometry.\\
That we are allowed to use the above procedure in the case of NUT spacetime, is due to the fact that one can maximally extend the Taub-NUT metric  with a kruskal-like extension \cite{Miller}. This is discussed very briefly in the appendix.
Now going back to the surface $Y = Y (\zeta,\rho)$, its equation can be obtained by  equating the two metrics \eqref{n11}, and \eqref{Rindler} at a given time $t=\frac{\zeta}{\kappa}$, leading to
\begin{equation}
 d \rho^2 + dY^2 = \frac{dr^2}{f(r)},
\end{equation}
which upon taking into account \eqref{Trax1}, we get
\begin{equation}\label{Y}
Y=\int_{r_H}^r \sqrt{\frac{1}{f(r)}\left(1-\frac{1}{4 \kappa^2}\left(\frac{d f}{d r}\right)^2\right)} d r.
\end{equation}
Now for the NUT spacetime with $\kappa =\frac{1}{2 \left(\sqrt{l^2+m^2}+m\right)}$, the above equation together with equations \eqref{Trax}-\eqref{Trax1}comprise the following embedding equations for the NUT region $r_H <r < \infty$,
\begin{flalign}\label{ETrax}
\zeta &= \frac{t}{2 \left(\sqrt{l^2+m^2}+m\right)} \\
\rho &= 2 (\sqrt{l^2+m^2}+m)\sqrt{\frac{r^2-l^2-2 m r}{r^2+l^2}}\\
Y &= \int_{r_H}^r \sqrt{\frac{r^2+l^2}{r^2-2 m r-l^2}\left(1-\frac{4 \left(\sqrt{l^2+m^2}+m\right)^2 \left(l^2 m-2 l^2 r-m r^2\right)^2}{\left(l^2+r^2\right)^4}\right)} \, dr.
\label{Trax11}
\end{flalign}
Embedding equations for the other three regions of the extended radial NUT plane (refer to the appendix) can be obtained by simple symmetry arguments as in the case of Kruskal spacetime \cite{Marolf}.\\
Using  equation \eqref{Trax11} along with the coordinates $(T, X)$ obtained from equations \eqref{TX1}-\eqref{TX2}, and similar equations for the other three regions of the extended radial NUT, the  dynamic embedding diagrams for different NUT spacetimes are obtained as shown in Figs. $12-14$.
%%%%%%%%%%%%%%%%%%%%%%%%%%%%%%%%%%
\begin{figure}\label{fig PNUT}
\includegraphics[scale=1.1]{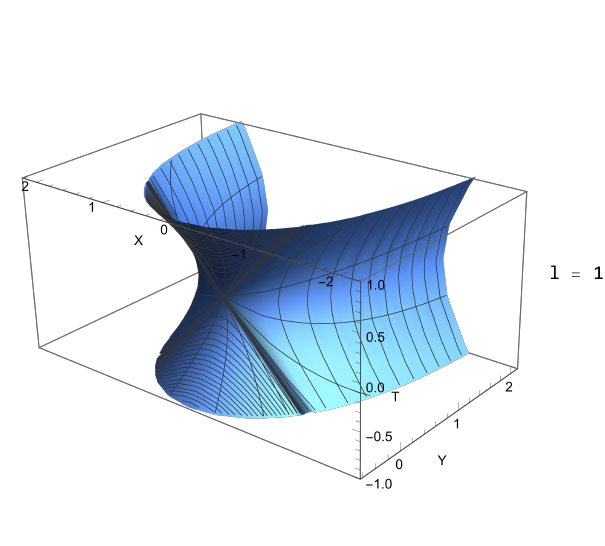}
\caption{Dynamic embedding diagram for pure NUT spacetime with $m=0$ and $l=1$}
\end{figure}
%%%%%%%%%%%%%%%%%%%%
\begin{figure}\label{fig PNUTvsSch}
\includegraphics[scale=1.1]{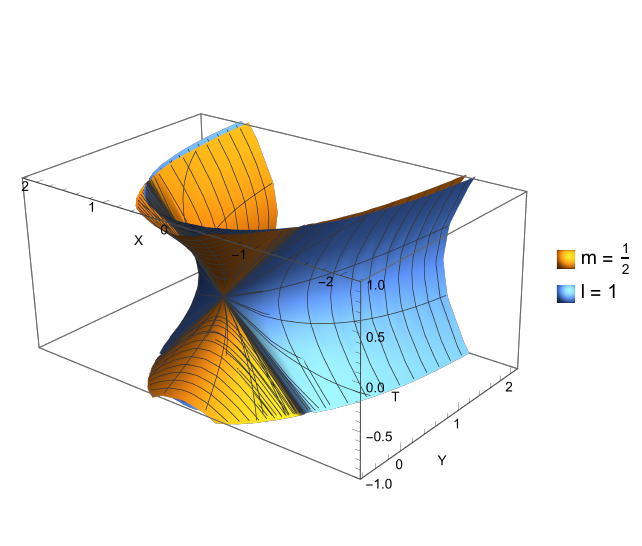}
\caption{Dynamic embedding diagrams for the Pure NUT ($l=1$) and Schwarzschild ($m=1/2$) spacetimes.}
\end{figure}
%%%%%%%%%%%%%%%%%%%%%%%%%%%%%%%%%%%%%%%%%%%%%%%%%%%
\begin{figure}\label{fig NUTvsNUT}
\includegraphics[scale=1.1]{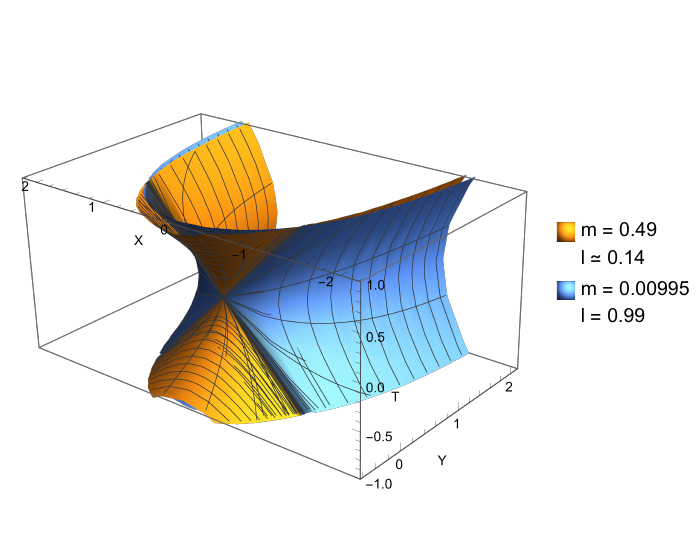}
\caption{Dynamic embedding diagrams for NUT spacetime with two different sets of values for its parameters.}
\end{figure}
%%%%%%%%%%%%%%%%%%%%%%%%%%%%%%%%%%%%
Embedding diagram of radial {\it pure} NUT spacetime ($m=0$) is shown in Fig. 12. The same embedding diagram for the two limiting cases of the NUT spacetime, namely pure NUT, and the Schwarzschild ($l=0$) spacetimes, with the same horizon at $r_H = 1$, are illustrated together in Fig. 13.
Embedding diagrams of NUT spacetimes  with two different sets of values for the mass and NUT parameters are shown together in Fig. 14.
In this case, to compare the effects of mass and NUT parameters, we have a NUT spacetimes with a small  mass parameter, and another with a small NUT parameter. \\
As expected, the null lines  $X=\pm T$ (corresponding to $r=r_H$, and equivalently to $Y=0$) in these diagrams divide the surface into four regions. Two of them, extended to positive values of $Y$, are related to the outside of the horizon (corresponding to the two ${\rm NUT}_+$ regions), while those with negative values of $Y$ are concerned with $r < r_H$ ( corresponding to the two Taub regions).
On these diagrams two different families of coordinate curves are drawn which are the $r = constant$ and $t = constant$ curves in terms of the original Schwarzschild-like coordinates in NUT spacetime metric \eqref{n1}.\\
Since the NUT spacetime is asymptotically flat, the  analysis of the information encoded in these diagrams is very similar to that of Kruskal spacetime, except the facts that the interior region is not a black hole region but a Taub region, and the $r=0$ is not a singularity but a regular point.
%%%%%%%%%%%%%%%%%%%%%%%%%%%%%%%%%%%
\section{Conclusions}
We have obtained both the spatial and dynamic embedding diagrams for some stationary spacetimes including NUT and pure NUT spacetimes, and discussed their characteristics in comparison with the corresponding embedding diagrams in Schwarzschild spacetime. In the case of spatial embedding diagrams the effect of different parameters were discussed by comparing their Gaussian and mean curvatures. One should be very careful with drawing conclusions from either of these embedding diagrams about the intrinsic properties of the corresponding spacetimes and their effects on particle trajectories, specially those related to spacetime curvature. For example, the geometry of the  Flamm's paraboloid could help us to calculate the contribution of the {\it spatial} geometry on the precession of a planet's perihelion, as well as on the light bending \cite{Rindler}.\\
On the other hand, the inclusion of the time coordinate provides the spacetime embedding diagrams with a dynamic nature so that they include  more visually traceable features of radial particle trajectories in the corresponding spacetime. From the relative behaviour of these trajectories, embedded in $1+2$-dimensional Minkowski spacetime, one could obtain physical effects rooted in the curved nature of the underlying spacetime. These include gravitational redshift and gravitational tidal effects near the horizon  and far away from it in the asymptotically flat region of the spacetime \cite{Marolf}.
%%%%%%%%%%%%%%%%%%%%%%%%%%%%%%%%%%%%%%%%%%%%%%
\section *{Acknowledgments}
The authors would like to thank University of Tehran for supporting this project under the grants provided by the research council. This work is based upon research funded by the Iran national science foundation (INSF) under the project No. 4005058.
%%%%%%%%%%%%%%%%%%%%%%%%%%%%%%%%%%%%%%%%%%%%%%%%%%%%%%
\section*{APPENDIX}
In the maximal extension of NUT spacetime to cover $-\infty < r < \infty $, there are coordinate singularities where $f(r) =0$ at $r_{\pm} = m  \pm (m^2 + l^2)^{1/2}$ ($r_+ \equiv r_H$) dividing the extended spacetime into three distinct regions. These are  the $r > r_+$ region (NUT$_+$), the $r < r_- $ region (NUT$_-$)  both with $f(r) > 0$, which are separated by the third region, the so called Taub region where $r_- <r < r_+$ in which $f(r) < 0$ \cite{Miller, PG}.\\
For our purpose we confine the extension to $0 < r < \infty $,
which divides this region to two NUT$_+$ regions ($r > r_+$)  and two half-Taub regions ($0 <r < r_+$). This is achieved by introducing the following Kruskal-type coordinates;
\begin{flalign}
\tilde T = (\frac{r- r_+}{r_+})^{\frac{1}{2}}(\frac{r- r_-}{r_-})^{\frac{r_-}{2r_+}} e^{r/2r_+} \sinh(\frac{t}{2r_+})   \\
\tilde X = (\frac{r- r_+}{r_+})^{\frac{1}{2}}(\frac{r- r_-}{r_-})^{\frac{r_-}{2r_+}} e^{r/2r_+} \cosh(\frac{t}{2r_+})
\end{flalign}
or conversely,
\begin{flalign}
(\frac{r- r_+}{r_+})(\frac{r- r_-}{r_-})^{r_-/r_+} e^{r/r_+} = \tilde{X}^2 -{\tilde T}^2   \\
t=2 r_+ \tanh^{-1} ({\tilde T}/\tilde{X})
\end{flalign}
under which  the NUT radial plane \eqref{n11} transforms into,
\begin{flalign}
ds^2 = 4\frac{r_-{r_+}^3}{r^2 + l^2} (\frac{r- r_-}{r_-})^{1- \frac{r_-}{r_+}} e^{-r/r_+} ( d\tilde{T}^2 - d{\tilde X}^2)
\end{flalign}
in which the horizon $r=r_+$ now corresponds to null lines $\tilde T=\pm {\tilde X}$ in Kruskal-type coordinates, and the other $r=constant$ hypersurfaces, including $r=0$, are hyperbolas. Our embedding diagram in the text is confined to the hatched region of the Penrose diagram of the full extended Taub-NUT solution in Fig. 15.
%%%%%%%%%%%%%%%%%%%%%%%%%%%%%%%%%%%%%%%%
\begin{figure}\label{TN}
\includegraphics[scale=0.60]{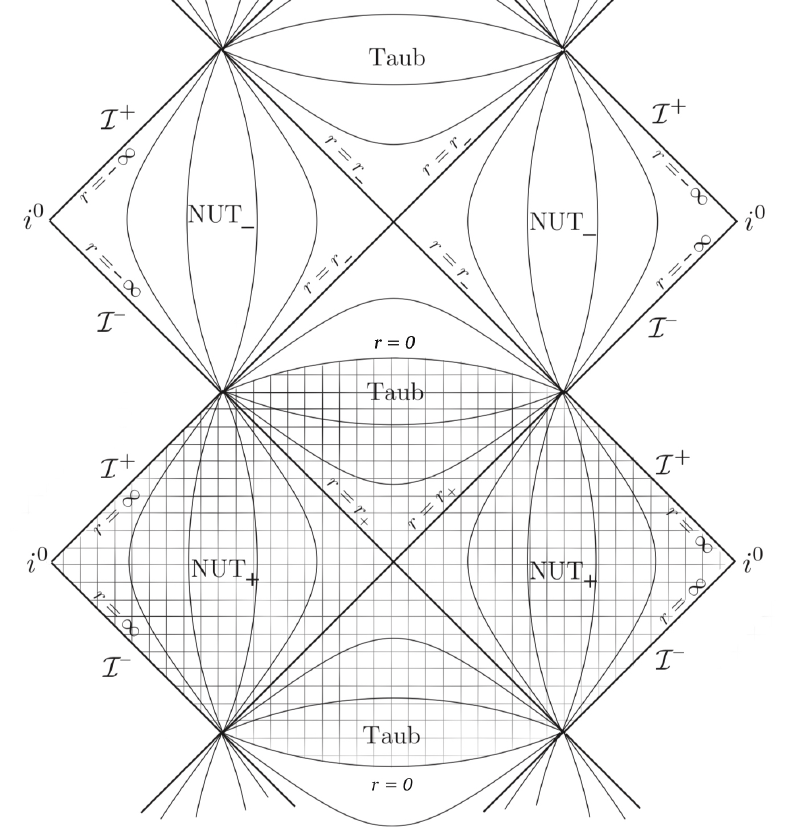}
\caption{Penrose diagram of maximally extended Taub-NUT solution. The hatched region covers $0 < r < \infty$.}
\end{figure}
%%%%%%%%%%%%%%%%%%%%%%%%%%%%%%%5
Its resemblance to the well-known Kruskal spacetime (eternal Schwarzschild black hole) is obvious.\\
%%%%%%%%%%%%%%%%%%%%%%%%%%%%%%%%%%%%%%%%%%%%%%%%

\end{document}